\documentclass[conference]{IEEEtran}
\IEEEoverridecommandlockouts
\usepackage[numbers,comma,sort&compress]{natbib}
\usepackage{hyperref}
\usepackage{amsmath,amssymb,amsfonts}
\usepackage{algorithmic}
\usepackage{graphicx}
\usepackage{booktabs}
\usepackage{textcomp}
\usepackage{xcolor}
\usepackage{soul}
\usepackage{tikz}
\usepackage{xspace}
\usepackage{subcaption}
\usepackage[font=small]{caption}
\usepackage{flushend}

\definecolor{gpuorange}{RGB}{235,125,10}
\definecolor{cpublue}{RGB}{30,90,235}
\newcommand{\gpu}[1]{\textcolor{gpuorange}{#1}}
\newcommand{\cpu}[1]{\textcolor{cpublue}{#1}}
\def\BibTeX{{\rm B\kern-.05em{\sc i\kern-.025em b}\kern-.08em
    T\kern-.1667em\lower.7ex\hbox{E}\kern-.125emX}}
\begin{document}

\newcommand{\name}{MIRC\xspace}

\title{Multi-scale Image Representation Compression}

\author{%
\textbf{Tianhao Peng}$^{1}$,
\textbf{Ho~Man~Kwan}$^{1}$,
\textbf{Fan~Zhang}$^{1}$, 
\textbf{Shan Liu}$^{2}$,
\textbf{David~Bull}$^{1}$ \\
$^{1}$Visual Information Lab, University of Bristol, UK \\
$^{2}$Tencent Media Lab, Palo Alto, USA \\
\texttt{\{tianhao.peng, hm.kwan, fan.zhang, dave.bull\}@bristol.ac.uk} \\
\texttt{shanl@global.tencent.com}
}
\maketitle

\begin{abstract}

Overfitted codecs have demonstrated promising performance for image and video compression. In particular, for image compression, the Cool-chic family of models has shown competitive performance against scene-agnostic models, with orders of magnitude lower decoding complexity at the cost of a longer overfitting process. However, these overfitted image codecs are not fully optimized toward the rate-distortion objective: their network weights remain in full precision during training, and the associated quantization parameters are selected in a separate post-training stage. Furthermore, their synthesis operates at a single scale, which overlooks cross-scale redundancy. In this paper, we propose \name, an overfitted image codec in which every coded component, including the latents, the synthesis network, and the entropy models, is quantized and entropy coded under a single rate-distortion objective, adopting the end-to-end compression pipeline of the neural video representation codec NVRC. We further introduce a multi-scale representation with cross-stage parameter sharing, which improves coding efficiency at a small transmitted overhead. On the CLIC2020 professional validation set, \name achieves a $10.5\%$ BD-rate saving against VVC (VTM 22.0). Moreover, \name offers a family of configurations spanning 1.2 to 2.9 kMAC per pixel, so the decoding budget can be selected to match the deployment target.

\end{abstract}


\section{Introduction}
Recent advances in neural image compression have delivered promising results, with the latest neural image codecs \cite{jiang2025mlic++} outperforming the best conventional codecs. In contrast to conventional codecs, which rely on expert-designed coding pipelines, neural codecs learn directly from data and have advanced rapidly through end-to-end optimization \cite{bull2021intelligent}, hyperprior \cite{balle2018variational}, other more advanced priors \cite{cheng2020learned}, and network architecture enhancements \cite{he2022elic, jiang2023mlic, gao2026advances}. However, although these neural image codecs have pushed the boundaries of compression performance, their computational complexity is prohibitively high, making them challenging to deploy on typical consumer devices and thus limiting their applicability.

\begin{figure}
    \centering
    \includegraphics[width=1\linewidth]{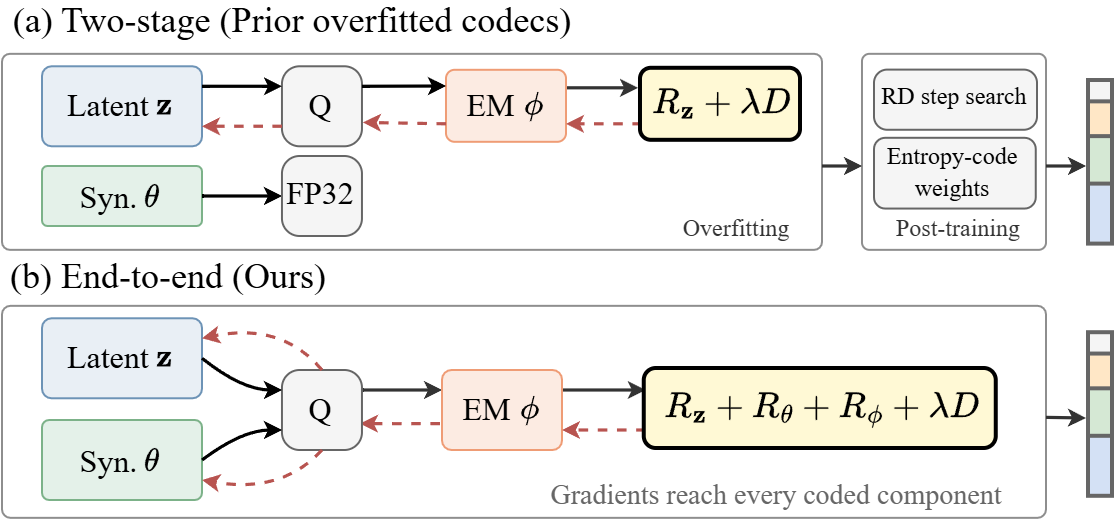}
    \caption{\textbf{(a)} Prior overfitted codecs \cite{ladune2023cool, coolchic_repo, kim2024c3} train their latents against a
    rate-distortion objective, but quantize and entropy-code the network
    weights in a separate stage. \textbf{(b)} \name (ours)  optimizes
    every coded component, including latents, synthesis weights
    and entropy model (EM) weights, under a single rate-distortion objective with
    quantization-aware training.}
    \label{fig:concept}
    \vspace{-10pt}
\end{figure}

More recently, lightweight image codecs based on overfitted representations have been proposed \cite{dupont2021coin, dupont2022coin++}. These codecs \textit{overfit} a specific image instance with a compact network rather than optimizing a model that must generalize across a diverse image distribution. Notably, the Cool-chic family of models \cite{leguay2024cool, ladune2026cool} has demonstrated promising performance, with the latest variants closely approaching strong, high-complexity neural image codec baselines, and outperforming state-of-the-art conventional codecs. Despite this promising performance, these models still lag behind the strongest generalized neural codecs, and their encoding time remains substantial due to the overfitting process.

In this paper, we address the aforementioned performance gap by revisiting the optimization pipeline used in existing overfitted image codecs. As shown in Fig.~\ref{fig:concept}. (a), these methods typically first overfit a representation network to an image, and then perform post-training hyperparameter selection to determine the quantization parameters that best trade off rate and distortion. Although effective, in particular when the decoder is extremely compact, this two-stage pipeline is inherently sub-optimal: the representation and quantization parameters are not jointly optimized with the same rate-distortion objective, which limits the representation capacity and the overall compression performance. In contrast, recent overfitted video codecs have adopted a more advanced end-to-end optimization pipeline \cite{kwan2024nvrc,kwan2026enhanced, gao2025givic}, where the representation network, entropy model, and  quantization parameters are jointly trained with a single rate-distortion loss, leading to promising compression performance.

We therefore propose \name, an overfitted neural image codec that extends the NVRC framework \cite{kwan2024nvrc} for image compression. It adopts a hierarchical compression and quantization pipeline, in which the image representation is optimized alongside the entropy model and quantization parameters. This avoids the sub-optimal post-training hyperparameter selection stage and enables the representation to be learned directly under the target coding objective. However, directly applying the original NVRC representation, i.e., HiNeRV \cite{kwan2023hinerv}, to image coding is inefficient, as its large parameter is amortized effectively over video frames. To address this, we introduce a multi-scale representation with parameter sharing, defined to retain sufficient representational capacity of HiNeRV while reducing the overhead for single-image coding. 

The proposed multi-scale image representation compression framework has been fully evaluated on two widely used test sets: Kodak and CLIC2020 professional validation. The results show that \name has achieved a $10.5\%$ BD-rate saving against VVC VTM 22.0 (All Intra) on the CLIC2020 pro. and at least 5\% coding gains over other state-of-the-art overfitted image codecs, including Cool-chic 4.0. More importantly, while maintaining a relatively low decoding complexity.

\section{Method}

\begin{figure}
    \centering
    \includegraphics[width=0.95\linewidth]{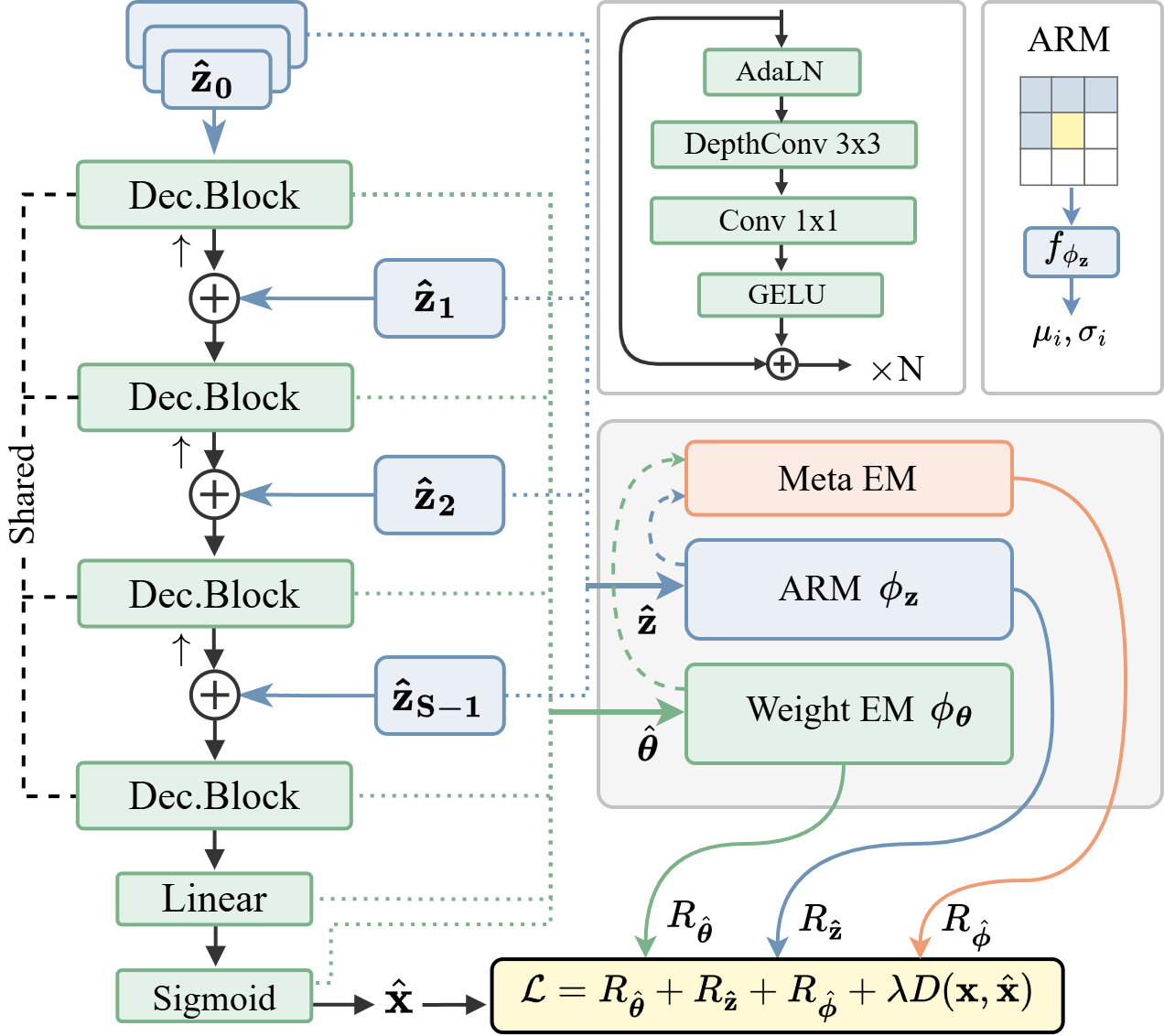}
    \caption{Overview of the proposed \name framework, with insets detailing the shared decoder block (top) and the causal context of the ARM (right). All coded components ($\hat{\mathbf{z}}$, $\hat{\boldsymbol{\theta}}$, $\hat{\boldsymbol{\phi}}$) are optimized under the single rate-distortion objective.}
    \label{fig:framework}
       \vspace{-10pt}
\end{figure}

As illustrated in Fig.~\ref{fig:framework}, the proposed \name represents an image $\mathbf{x} \in \mathbb{R}^{H \times W \times 3}$ using a set of parameters optimized for this image, which are subsequently compressed into the bitstream for transmission. The representation consists of multi-scale latent grids $\mathbf{z}$ and a lightweight synthesis network with weights $\boldsymbol{\theta}$, which decodes the latents into the reconstructed image. The latent grids are coded by an autoregressive entropy model (ARM) with parameters $\boldsymbol{\phi}_{\mathbf{z}}$, and the synthesis weights are coded by a non-parametric weight entropy model with parameters $\boldsymbol{\phi}_{\boldsymbol{\theta}}$. Both of them are in turn coded by a meta entropy model and included in the bitstream. The remainder of this section describes the compression framework, covering quantization, entropy coding, and end-to-end optimization, and the proposed multi-scale representation.

\subsection{Compression framework}
\label{subsec:compress_framework}

\noindent\textbf{Quantization.} Both parameter groups are quantized with uniform steps:
\begin{equation}
    \hat{\mathbf{z}} = \lfloor \mathbf{z} / \delta_{\mathbf{z}} \rceil
    \cdot \delta_{\mathbf{z}}, \qquad
    \hat{\boldsymbol{\theta}} = \lfloor \boldsymbol{\theta} /
    \delta_{\boldsymbol{\theta}} \rceil \cdot \delta_{\boldsymbol{\theta}},
    \label{eq:quant}
\end{equation}
where the latent step $\delta_{\mathbf{z}}$ is a single constant shared by all grids \cite{ladune2023cool, kim2024c3}, and the weight step $\delta_{\boldsymbol{\theta}}$ is configured separately for each parameter tensor and transmitted as side information within the bitstream. While a separate latent step could be learned for each grid, it adds little flexibility: scaling a grid at a fixed step has an equivalent effect to reducing the step size, so a learned step would increase the optimization space and introduce additional parameters. 

Instead, the weight step is computed adaptively as $\delta_{\boldsymbol{\theta}} = (\theta_{\max} - \theta_{\min}) / (2^b - 1)$, where $\theta_{\max}$ and $\theta_{\min}$ denote the maximum and minimum values \emph{within a tensor}, $b$ is a fixed bit depth ($b=8$), and all parameters in that tensor, e.g., those of a single convolutional layer, share the same step. This corresponds to fixed-width quantization with a per-tensor step: in practice, such moderate quantization typically incurs little performance loss, while keeping the rate-distortion trade-off easy to optimize, even when using the noisy quantization proxy \cite{agustsson2020universally, kim2024c3} during training. It also bounds the symbol alphabet for entropy coding, while the transmitted steps add only a negligible per-tensor overhead. Note that the average rate can still be lower than $b$ bits per parameter, since a learned entropy model, described below, is used to model and optimize the distribution of the quantized parameters.

\vspace{5pt}
\noindent\textbf{Entropy coding.} The quantized symbols in Eq.~\eqref{eq:quant} are then losslessly encoded into the bitstream, and the rate they consume is dependent on how accurately the distribution of each symbol is modeled. The two quantized parameter groups $\{\hat{\mathbf{z}}, \hat{\boldsymbol{\theta}}\}$ therefore require different models. The latents $\hat{\mathbf{z}}$ retain spatial correlation, following Cool-chic 4.0 \cite{coolchic_repo}, they are coded by the ARM, which factorizes their distribution over positions:
\begin{equation}
    p(\hat{\mathbf{z}}) = \prod_i p\big(\hat{z}_i \mid \mathbf{c}_i\big),
    \qquad
    (\mu_i, \sigma_i) = f_{\boldsymbol{\phi}_\mathbf{z}}(\mathbf{c}_i),
    \label{eq:arm}
\end{equation}
in which $\mathbf{c}_i$ collects the previously decoded neighbors of position $i$, the residual MLP $f_{\boldsymbol{\phi}_\mathbf{z}}$ maps this causal context to the mean and scale of a Gaussian, and $p(\hat{z}_i \mid \mathbf{c}_i)$ is the probability assigned by this Gaussian to the quantization bin of $\hat{z}_i$.

 It is noted that the synthesis network parameters $\hat{\boldsymbol{\theta}}$ do not exhibit a comparable spatial structure to the latents. As a result, a context model like the ARM is not suitable here. While the latents can be freely adjusted to follow any imposed prior, with the synthesis network adapting accordingly, the weights, in contrast, parameterize the decoder itself, so their distribution is determined solely by fitting the image and cannot be assumed to take any particular form. We therefore code the weights with a non-parametric model that can fit an arbitrary distribution \cite{balle2018variational}, referred to as the weight entropy model $\boldsymbol{\phi}_{\boldsymbol{\theta}}$. Since such a model is associated with a relatively large number of parameters, we share a single model across all synthesis layers rather than learning one per layer. Thus, the transmitted overhead does not grow with the network. This leaves the entropy model parameters themselves, namely the ARM parameters $\boldsymbol{\phi}_{\mathbf{z}}$ and the weight entropy model parameters $\boldsymbol{\phi}_{\boldsymbol{\theta}}$. Following the hierarchical parameter coding of NVRC \cite{kwan2024nvrc}, both are coded by a meta entropy model, which applies another shared non-parametric model to $\boldsymbol{\phi}_{\mathbf{z}}$ and a Gaussian model to $\boldsymbol{\phi}_{\boldsymbol{\theta}}$. The small number of parameters of the meta entropy model is stored with a uniform model, closing the hierarchy at negligible cost.

\vspace{5pt}
\noindent\textbf{End-to-end optimization.} We optimize all of these
parameters jointly under a single rate-distortion objective:
\begin{equation}
    \mathcal{L} = R_{\mathbf{\hat{\theta}}} + R_{\mathbf{\hat{\phi}}}+ R_{\mathbf{\hat{z}}}+\lambda D(\mathbf{x},\hat{\mathbf{x}}),
    \label{eq:objective}
\end{equation}
where the rate terms are the negative log-likelihoods under the entropy
models above, e.g., $R_{\hat{\mathbf{z}}} = -\sum_i \log_2 p(\hat{z}_i \mid
\mathbf{c}_i)$. $D$ is the distortion between $\mathbf{x}$ and the
reconstruction $\hat{\mathbf{x}}$, e.g., the mean squared error, and
$\lambda$ controls the trade-off between rate and quality. The only
non-differentiable operation is the rounding in Eq.~\eqref{eq:quant}, which
is approximated during training through soft-rounding with additive
Gaussian noise \cite{agustsson2020universally, kim2024c3}.
  
With this proxy in place, the gradients of $\mathcal{L}$ propagate to every coded
component, including the weights of the entropy model itself. This is the
key difference between our work and the Cool-chic family
(Fig.~\ref{fig:concept}): while these codecs also train their latents
with a rate term, the network weights typically remain in full
precision during training, with their quantization parameters selected through
a post-training search \cite{coolchic_repo, kim2024c3,li2026moric}.

\subsection{Multi-scale representation}
\label{subsec:representation}
\noindent\textbf{Synthesis network.} The synthesis network reconstructs the image in a multi-scale manner through $S$ stages of network blocks, as shown in Fig.~\ref{fig:framework}. In the $s$-th stage, the features from the previous stage, $s-1$, are first upsampled by a factor of two, except for the first stage $s=0$. A feature encoding is then extracted from the quantized latent grid $\hat{\mathbf{z}}_{s}$, and projected into affine parameters that modulate the upsampled features (reducing to a bias for $s=0$) \cite{gao2025pnvc, kwan2026enhanced}. The modulated features are then processed by $N$ decoder blocks. Following HiNeRV \cite{kwan2023hinerv}, we use bilinear interpolation and hierarchical encoding for upsampling, whose high parameter efficiency is important because these parameters must also be encoded into the bitstream.

\vspace{5pt}\noindent\textbf{Multi-scale latents and decoder features.} Following recent works on overfitted codecs \cite{kim2024c3}, the latents are organized to match the decoder scales. At the first stage, an $L$-level pyramid $\hat{\mathbf{z}}_0$ forms the input \cite{kwan2023hinerv}, while each subsequent upsampling is paired with an additional single-level grid $\hat{\mathbf{z}}_s$ \cite{ladune2023cool, kim2024c3, kwan2025ultra, kwan2026enhanced}. The finest level of $\hat{\mathbf{z}}_0$ and each single-level grid $\hat{\mathbf{z}}_s$ are stored at the same resolution as the corresponding decoder features, while the remaining $L-1$ levels of $\hat{\mathbf{z}}_0$ are stored at lower resolutions, following \cite{kwan2023hinerv}. The latent grids modulate the decoder features in a FiLM-like manner \cite{perez2018film, gao2025pnvc}, as
\begin{align}
     [\,\boldsymbol{\gamma}_{s},\ \boldsymbol{\beta}_{s}\,] &= P_{s}\big(\mathrm{Interp}(\hat{\mathbf{z}}_{s})\big), \label{eq:mod}
\end{align}
\vspace{-15pt}
\begin{align}
    \mathbf{X}_{s} &= F_{s}\big(\boldsymbol{\gamma}_{s} \odot U(\mathbf{X}_{s-1}) + \boldsymbol{\beta}_{s}\big),  \qquad 0 \le s < S-1, \label{eq:decoder}
\end{align}
in which $U$ denotes bilinear upsampling, $F_{s}$ denotes the decoder blocks at stage $s$, $P_{s}$ is a linear projection that predicts the per-stage scale and shift, and $\mathrm{Interp}(\cdot)$ denotes the interpolation used to sample the latent grids at the decoder resolution. We define $U(\mathbf{X}_{-1}) \equiv \mathbf{0}$, so that stage $0$ reduces to $\mathbf{X}_0 = F_0(\boldsymbol{\beta}_0)$, i.e., the block is driven purely by the predicted shift. The hierarchical grid encoding mechanism from \cite{kwan2023hinerv} is omitted from Eq.~\ref{eq:decoder} for simplicity.

\vspace{5pt}\noindent\textbf{Decoder blocks.} Each decoder block uses a depthwise-separable convolution \cite{howard2017mobilenets}, i.e., a depthwise convolution followed by a pointwise linear layer with a GELU activation \cite{hendrycks2016bridging}, and a skip connection \cite{he2016deep} around the block. After the last stage, a linear layer followed by a sigmoid activation maps the latent features to image-domain pixels to reconstruct $\hat{x}$.

\vspace{5pt}\noindent\textbf{Parameter sharing between stages.} While the multi-scale decoder better captures cross-scale redundancy, it introduces additional parameters that must be coded in the overfitted setting, which may harm the rate-distortion trade-off. This overhead weighs particularly heavily for a single image, which offers far fewer pixels than a video to amortize the parameter cost \cite{kwan2023hinerv}.  To mitigate this, we optionally share parameters across decoder stages, i.e., the $n$-th block is tied across all stages. The codec can then either reuse a shared block across scales or, through training, bias the shared parameters toward whichever scale contributes most to the loss. To further improve this sharing, we apply AdaLN \cite{peebles2023scalable}: a per-stage learned vector for scaling and shifting is applied within each shared block, allowing the shared block to retain stage-specific behavior at each stage.


\section{Experiments}
\label{subsec:experiments}
\subsection{Experimental setup}
\label{subsec:setup}

\noindent\textbf{Datasets and metrics.}
We evaluate the proposed image codec on two commonly used datasets, the Kodak dataset  \cite{kodak1993} and the CLIC2020 professional validation
dataset \cite{toderici2020clic}. The former contains 24 images with a size of
$768 \times 512$, while the CLIC 2020 pro. dataset comprises 41 images with resolutions ranging from
$384 \times 512$ to $1370 \times 2048$. Reconstruction quality is measured
by PSNR in the RGB space.

\vspace{5pt}\noindent\textbf{Baselines.}
\name is compared with three groups of codecs. The conventional anchor
is VVC VTM 22.0 \cite{bross2021overview} in the all-intra
configuration with QP $\in \{17, 22, \ldots, 42\}$. The
overfitted codecs include Cool-chic 4.0 \cite{coolchic_repo}, C3 \cite{kim2024c3} and
MoRIC \cite{li2026moric}. The autoencoder
baseline is MLIC++ \cite{jiang2025mlic++}. For VTM, Cool-chic 4.0, MLIC++, and our method, the reported rates are the sizes of the real bitstreams. Since C3 and MoRIC are not integrated with an entropy coder, their rates are cross-entropy estimates.

\begin{figure*}
    \centering
        \begin{subfigure}[b]{0.32\linewidth}
    \centering
    \includegraphics[width=\linewidth]{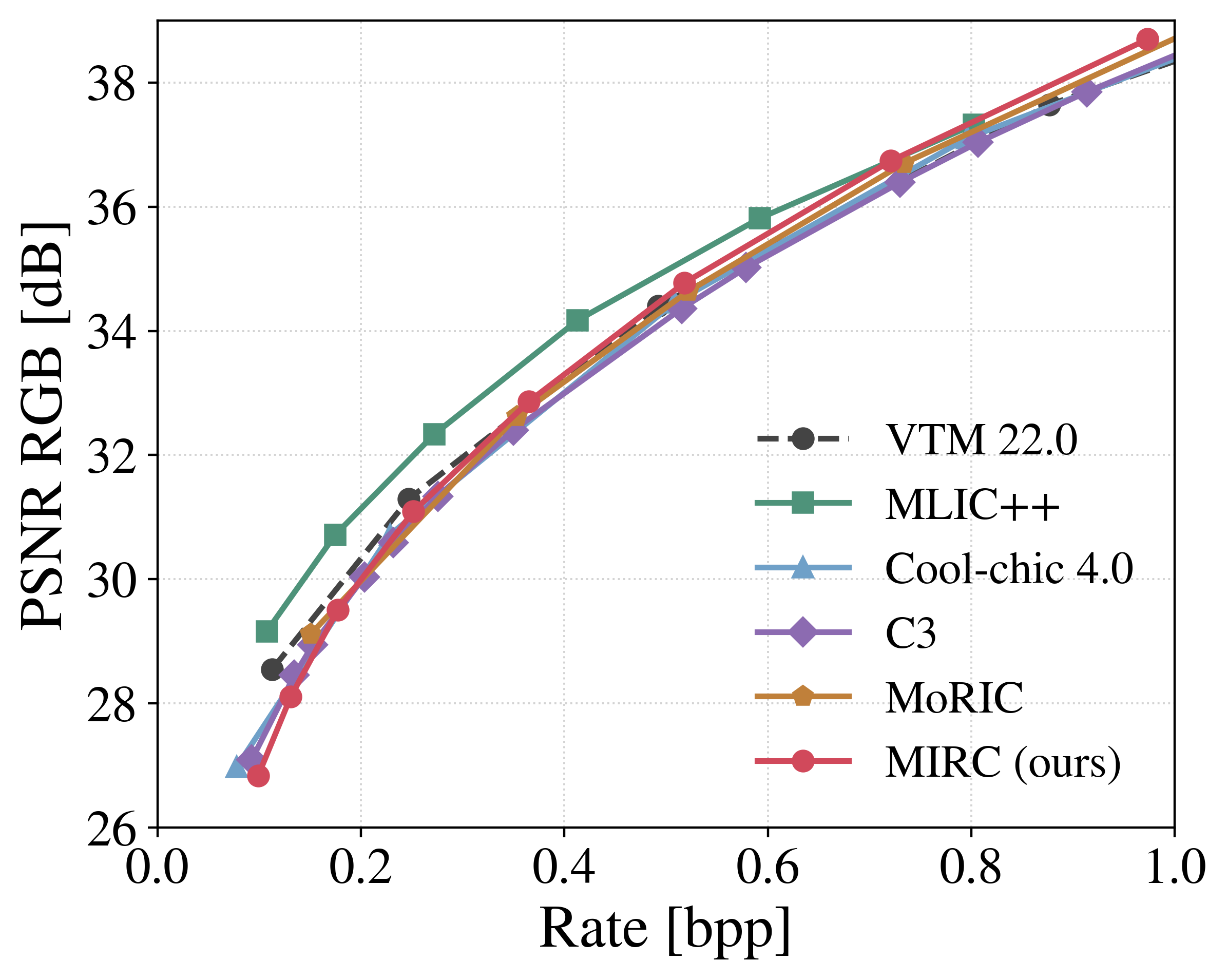} 
    \caption{Kodak.}
    \label{fig:rd_curve_kodak}    
  \end{subfigure}
      \begin{subfigure}[b]{0.32\linewidth}
    \centering
    \includegraphics[width=\linewidth]{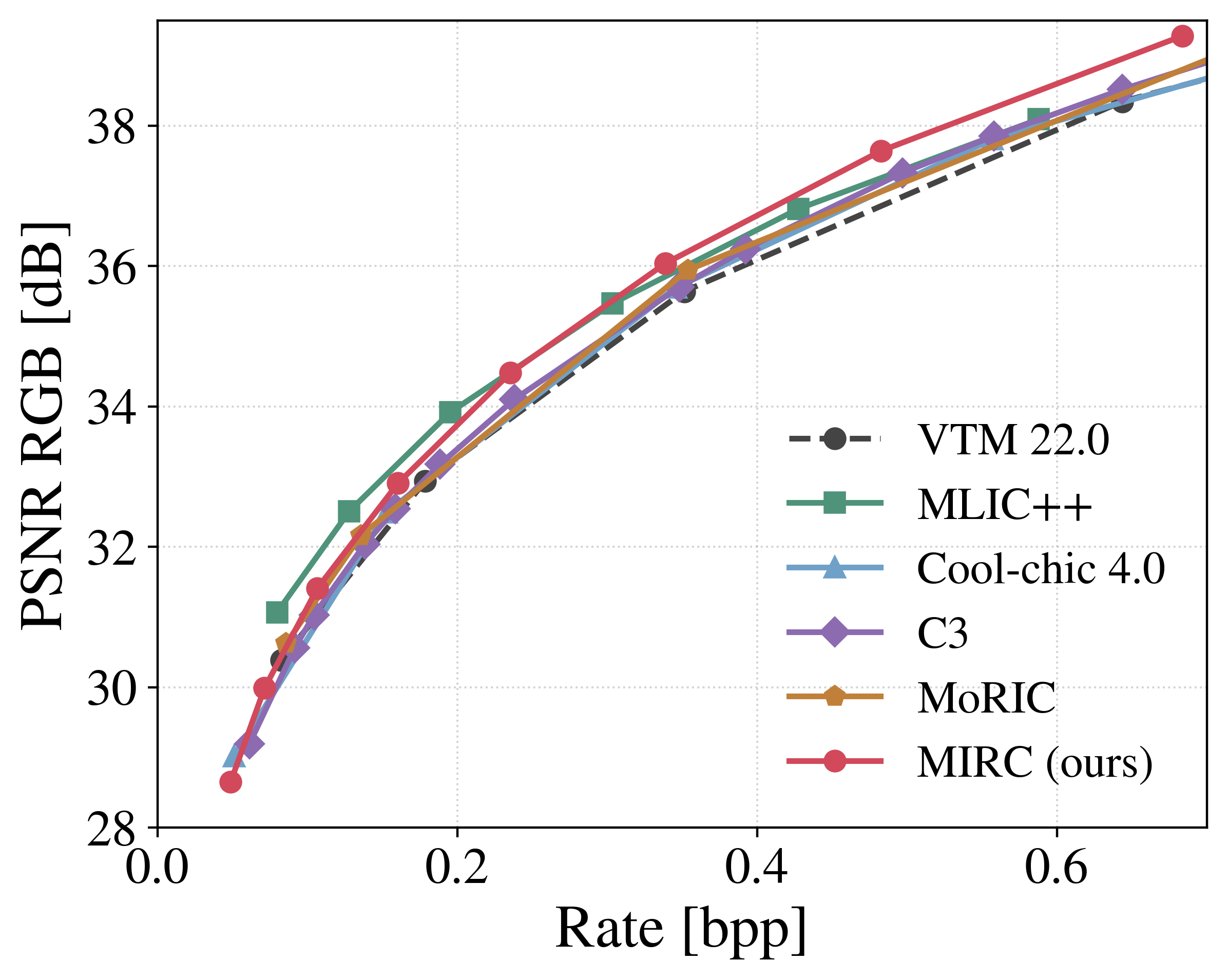} 
    \caption{CLIC2020 pro.}
    \label{fig:rd_curve_clic}    
  \end{subfigure} 
      \begin{subfigure}[b]{0.32\linewidth}
    \centering
    \includegraphics[width=\linewidth]{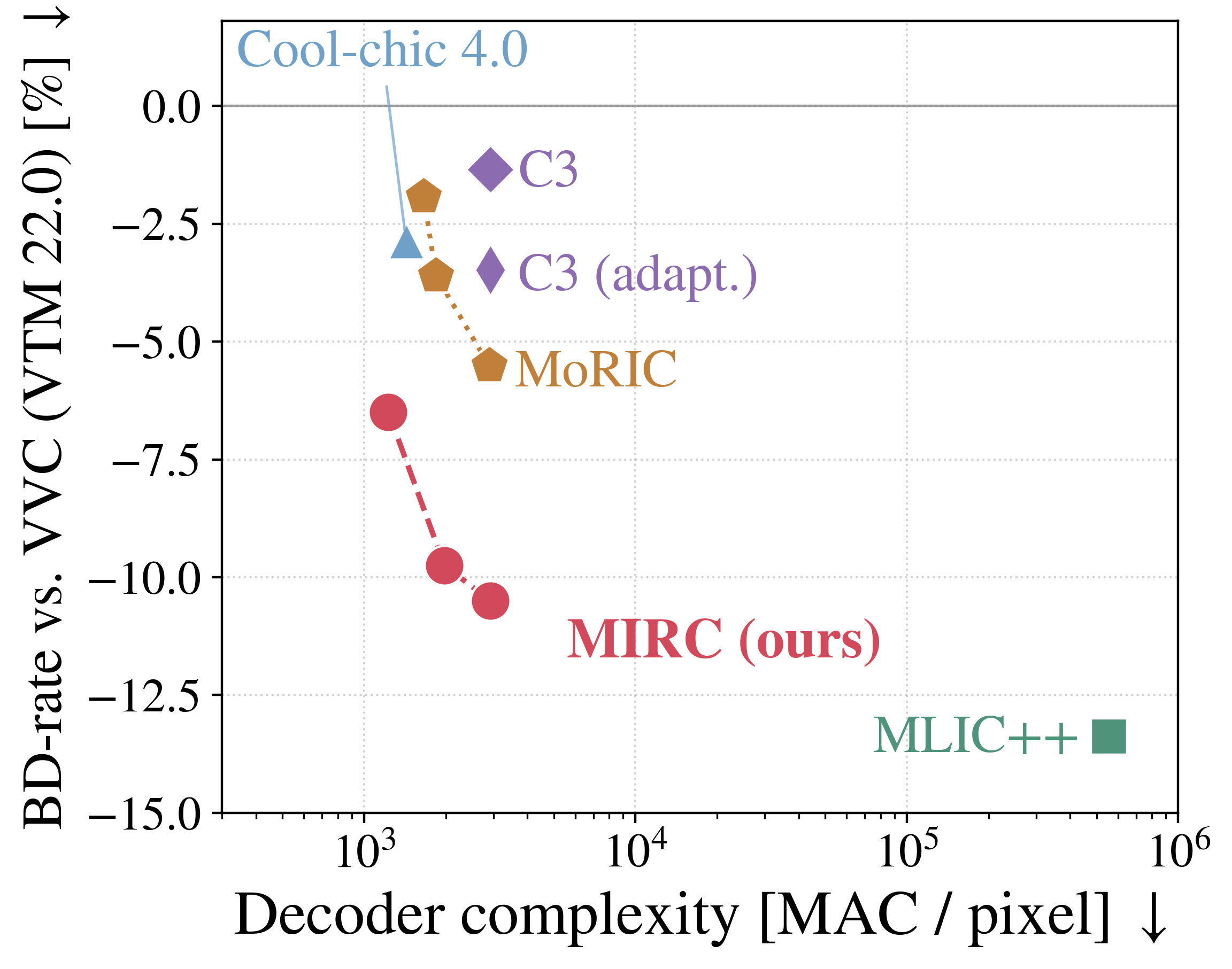} 
    \caption{CLIC2020 pro.}
    \label{fig:complexity} 
    
  \end{subfigure}
  \vspace{-5pt}
    \caption{Rate-distortion and complexity comparison. (a), (b): rate-distortion curves (PSNR versus bpp), with VTM 22.0 as the anchor and , on Kodak and the CLIC2020 professional validation dataset, respectively. (c): BD-rate against VTM 22.0 versus decoding complexity, in
multiply-accumulate operations (MAC) per pixel.}
\vspace{-10pt}
\end{figure*}
\vspace{5pt}\noindent\textbf{Implementation details.}
The default configuration of \name uses a synthesis network with $S=4$ stages, $N=2$ blocks per stage, and $L=3$ levels for the latent grids at the first stage, i.e., six sets of latent grids in total. The number of channels per grid is set to $\{4, 4, 4, 2, 1\}$ (the channels are padded so that the autoregressive model can be shared between different sets of grids). The kernel size of the synthesis network is set to 3. To reach different complexity levels, we vary the number of channels of the synthesis transform $C_{\mathrm{syn}}$ and the autoregressive model $C_{\mathrm{arm}}$, defining three complexity levels: \textit{Low} ($C_{\mathrm{syn}}=8$, $C_{\mathrm{arm}}=8$), \textit{Mid} ($C_{\mathrm{syn}}=12$, $C_{\mathrm{arm}}=12$), and \textit{High} ($C_{\mathrm{syn}}=16$, $C_{\mathrm{arm}}=16$). The shared decoder block is used only for the \textit{High} configuration.

To encode an image, \name is optimized for 100K iterations with mixed-precision training. All parameters are trained with Adam \cite{kingma2014adam} under a cosine learning-rate schedule from $10^{-2}$ to $10^{-6}$, and the quantization proxy follows the linear temperature and noise schedules described in Sec.~\ref{subsec:compress_framework}. Each complexity level is encoded at eight rate points by setting $\lambda \in$ {$\{32, 64, 128, 256, 512, 1024, 2048, 4096\}$. Further training details will be provided with the open-source implementation.

\subsection{Rate-distortion performance}
\label{subsec:rd_results}
Fig.~\ref{fig:rd_curve_kodak} and Fig.~\ref{fig:rd_curve_clic} show the rate-distortion curves of evaluated image codecs, while Fig.~\ref{fig:complexity} plots the trade off between BD-rate results and decoding complexity on the CLIC2020 dataset. At its largest configuration, \name achieves a BD-rate of $-10.5\%$ against VTM 22.0 and outperforms all other overfitted codecs by at least $5\%$. The autoencoder baseline, MLIC++, leads overall, but with  roughly 190 times the decoding complexity of our model. Across our three configurations, \name degrades gracefully as the decoding budget shrinks: moving from 2.9 to 2.0 kMAC per pixel costs only $0.8\%$ in BD-rate, and the smallest configuration still reaches $-6.5\%$ at 1.2 kMAC per pixel. Notably, this smallest variant already outperforms every configuration of the other overfitted codecs, at around half of their decoding complexity.

\begin{table}[!t]
    \centering
    \caption{Encoding and decoding time with complexity, measured and averaged on CLIC2020 pro. dataset using an NVIDIA RTX 4090 GPU and Intel Core i7-14700 CPU. Orange indicates \gpu{GPU}
computation, blue indicates \cpu{CPU} computation.}
    \scriptsize
    \label{tab:main}
    \begin{tabular}{rccc}
        \toprule
         & Encoding & \multicolumn{2}{c}{Decoding} \\
        \cmidrule(lr){2-2} \cmidrule(lr){3-4}
         & Time [s/image] & Complexity [kMAC/px] & Time [ms/Mpx] \\
        \midrule
        Cool-chic 4.0             & \gpu{3782}       & 1.43        & \cpu{135.3}     \\
        C3             & \gpu{2382}         & 1.79         & -  \\
        MoRIC (HOP) & \gpu{16,231}   & 2.93         & - \\
        \midrule
        \name (\textit{High})               & \gpu{3615}     & 2.92    & \gpu{17.8} \\
        \bottomrule
    \end{tabular}
       \vspace{-10pt}
\end{table}

\begin{table}[!t]
    \centering
    \caption{Ablation of the main design choices. BD-rate is computed
    relative to the default \name configuration (\textit{High}) on ten images of CLIC2020 pro. dataset.}
    \label{tab:ablation}
    \small
    \begin{tabular}{lr}
        \toprule
        Configuration & BD-rate [\%] $\downarrow$ \\
        \midrule
        w/o Multi-scale decoder & +2.85 \\
        w/o Decoder block sharing between stages & +4.95 \\
        w/o End-to-end optimization & +13.01 \\
        \midrule
        \name (default) & 0.0 \\
        \bottomrule
    \end{tabular}
    \vspace{-10pt}
\end{table}

\subsection{Ablation study}
\label{subsec:ablation}

We compare the proposed MIRC against three ablated variants: (1) a single-scale decoder ($S=1$); (2) a multi-scale decoder whose blocks are not shared between stages; and (3) a variant whose network-parameter entropy model (weight/meta entropy models) is not learned and is optimized without the quantization proxy. All three components improve compression performance.

Removing end-to-end optimization hurts most ($+13.01\%$), showing that optimizing the parameter distribution jointly with the quantization proxy is critical. The multi-scale decoder and block sharing also help, but their interaction is telling: multi-scale \emph{without} sharing ($+4.95\%$) is worse than single-scale ($+2.85\%$). Adding scales thus pays off only once blocks are shared, which keeps the per-stage parameter overhead low enough to be worthwhile in this overfitted setting.

\section{Conclusion}
\label{subsec:conclusion}
This paper presents \name, an overfitted image codec that represents an image with multi-scale latent grids and a lightweight multi-stage decoder, and optimizes every coded component jointly under a single training objective. Experiments show that our proposed codec attains rate-distortion performance on par with the strongest overfitted image codec, while its shared multi-scale decoder spans a range of decoding budgets. At matched decoding complexity, it also outperforms all other overfitted codecs by a clear margin, without any post-training tuning of the coding pipeline. Similar to other overfitted codecs, the encoding of our method remains costly, so accelerating the overfitting process is the main direction for future work.

\small
\setlength{\bibsep}{5pt}
\bibliographystyle{ieeetr}
\bibliography{ref}

\end{document}